\documentstyle[seceq,epsf,mbf,wrapft]{ptptex}
\def\Ga    {g_{_{A}}}

\def\DS    {\Delta \Sigma}

\def\Du    {\Delta u_{\rm con}}
\def\Dd    {\Delta d_{\rm con}}

\pubinfo{}  %Editorial Office use

\preprintnumber[3cm]{%<-- [..]: optional width of preprint # column.
KUNS-1325\\PTPTeX ver.0.8\\ August, 1997}
\title{%        %You can use \\ for explicit line-break
Polarized structure functions from the lattice
}
\author{%       %Use \sc for the family name
T. {\sc Blum}\footnote{Talk given by T. Blum}\footnote{email: tblum@bnl.gov}
and S. {\sc Sasaki}\footnote{email: ssasaki@bnl.gov}
}

\inst{%         %Affiliation, neglected when [addenda] or [errata]
RIKEN BNL Research Center
\\
Brookhaven National Lab\\Upton, NY 11973-5000, USA
}

\recdate{%      %Editorial Office will fill in this.
February 17, 2000}

\abst{%         %this abstract is neglected when [addenda] or [errata]
We give a brief sketch of lattice structure function calculations
and review previous results for the axial coupling $g_A$.
We outline a new technique for treating fermions on the lattice
that preserves chiral symmetry, domain wall fermions.
Finally, we give preliminary results
for the nucleon spectrum using this new technique. Remarkably, a large
mass splitting between the $N$ and $N^*$, roughly 
consistent with experiment,
is produced in the calculation.
These results are encouraging for proposed calculations of
nucleon structure functions.
}

\begin{document}

\maketitle

%\tableofcontents

\makeatletter
\if 0\@prtstyle
\def\asp{.3em} \def\bsp{.26em}
\else
\def\asp{.3em} \def\bsp{.3em}
\fi \makeatother

\section{Introduction}

Structure functions are an 
important quantities for understanding
hadronic physics from first principles and for confronting experimental
measurements. In QCD, the
theory of the strong interaction, they probe the underlying
make-up of hadrons from the fundamental quarks and gluons.
A complete theoretical understanding of hadronic physics requires
calculations of these important functions in QCD (or equivalently, a set of
operator matrix elements). Because QCD is a confining theory,
the complete calculation of the structure functions is necessarily 
non-perturbative. A rigorous, systematic method for calculating
non-perturbatively in QCD is the lattice method, or lattice QCD.

The definition of nucleon structure functions follows 
from the factorization of the deep inelastic scattering (DIS) cross section
of electrons and nucleons.
\begin{eqnarray}
\sigma_{dis} &\sim& L_{\mu\nu} W^{\mu\nu},\\
\label{w def}
W^{\mu\nu}&=&\frac{1}{4\pi}\int {\rm d}^4x e^{iq\cdot x}\langle
P|[J^\mu(x),J^\nu(0)]|P\rangle
\end{eqnarray}
where $L_{\mu\nu}$ is the lepton tensor, $W_{\mu\nu}$ is the
hadronic tensor, and $J^\mu$ is the electromagnetic current. 
We are interested in the form of $W_{\mu\nu}$\cite{ROBERTS}.
\begin{eqnarray}
W^{\mu\nu}(p,q)&=&\left(-g^{\mu\nu}+\frac{q^\mu
q^\nu}{q^2}\right)F_1(x,Q^2)\nonumber\\
               &+&
          \frac{1}{p\cdot q}\left(p^{\mu}-\frac{(p\cdot q) q^\mu}{q^2}\right)
              \left(p^{\nu}-\frac{(p\cdot q) q^\nu}{q^2}\right)F_2(x,Q^2),    
\end{eqnarray}
where $F_1(x,Q^2)$ and $F_2(x,Q^2)$ are the {\it unpolarized} nucleon
structure functions. Similarly, for scattering off of polarized nucleons with 
spin $S$, we have
\begin{eqnarray}
W^A_{\mu\nu}(p,q,S)&=&i\epsilon_{\mu\nu\rho\sigma}\frac{Mq^\rho}{p\cdot q}
   \left(S^{\sigma}g_1(x,Q^2)+
   \left(S^\sigma-\frac{S\cdot q}{p\cdot q}\right)g_2(x,Q^2)\right),
\end{eqnarray}
where $g_1(x,Q^2)$ and $g_2(x,Q^2)$ are the {\it polarized} nucleon
structure functions. In the above, $Q^2=-q^2$ is the momentum transfer
from the electron to the struck nucleon, $p^\mu$ is the incoming nucleon 
momentum, and $x=Q^2/(2p\cdot q)$.

The structure functions $F_1$, $F_2$, $g_1$,
and $g_2$ are 
calculable from first principles in QCD. One simply needs to
calculate the matrix element in Eq.~\ref{w def}.

The nucleon state $|P\rangle$ is a low energy hadronic {\it bound} 
state of quarks
and gluons; thus the matrix element, Eq.~\ref{w def}, 
cannot be calculated in perturbation theory. We must use a
non-perturbative technique such as the lattice method to calculate it. However,
the lattice method works only in Euclidean space-time, while DIS occurs near
the light-cone in Minkowski space-time. 
Thus we cannot calculate the desired matrix element directly.
Instead, we resort to the operator product expansion; 
the non-local
operator $[J^\mu(x),J^\nu(0)]$ 
is given in terms of an infinite set of local operators.
The utility of the method is that most of these operators are 
suppressed by powers of $1/Q$. The organization of the expansion is
in terms of the so-called {\it twist} of the (local) operator, which
is defined as the dimension of the operator minus its spin, $T=d-s$.
The Wilson coefficients of the expansion, $C_n$ are calculable in perturbation
theory, and one can show from dimensional reasoning that 
$C_n\sim q^{-(T-2)}$, so at leading order we need only consider $T=2$.
We note that
the structure functions can be related to $T=2$ {\it distribution}
functions in the parton picture, {\it i.e.}, the probability of finding
a parton (quark or gluon) with momentum fraction $x$ of the nucleon
inside the nucleon. At tree level, the coefficient functions are simply
the charges of the partons.

In summary, the calculation of the matrix element is split into two
parts, a soft part given by the matrix elements of the local operators, and
a hard part given by the $C_n$. The dividing line, or scale $\mu$, 
where the split is made is arbitrary. However, at any finite order of
perturbation theory, there is residual scheme dependence in
$C_n$, so one cannot take $\mu$ too small.
In addition, present lattice simulations have a cut-off $1/a$
around 2-3 GeV. These considerations constrain the
range of $\mu$ values that can be used.

Looking in more detail, we see that the matrix elements calculated
on the lattice are actually moments (in $x$) of the structure functions,
or more precisely, their Mellin transforms. 
Thus in principle, to construct
the structure functions one must calculate all of the moments
and perform the inverse Mellin transform. Fortunately, for $x$ not
very close to 1, we only need the lowest few moments since rapidly, the higher
moments begin to probe only the high $x$ region.

The axial and tensor charges  are given by the
lowest moment of the distribution functions $g_1(x)$ and $h_1(x)$, the
probability of finding a parton with longitudinal or transverse polarization
inside a longitudinally or transversely polarized nucleon, respectively.
\begin{eqnarray}
      \int {\rm d}x\, g_1(x) &=& \Delta q,\\
      2 M_N S_\mu \Delta q 
      &=& \langle P,S|\bar q \gamma_\mu \gamma_5 q |P,S\rangle,\\
      \int {\rm d}x\, h_1(x) &=& \delta q, \\
      2 (S_\mu P_\nu - P_\mu S_\nu)\delta q
      &=& \langle P,S|\bar q i\sigma_{\mu\nu} \gamma_5 q |P,S\rangle,
\end{eqnarray}
where 
%-->  $\sum_{q=u,d,s}\Delta q$ 
$\DS=1/2\sum_{q=u,d,s}\Delta q$ 
may be interpreted as
the quark contribution to the nucleon spin.
Also, $\Delta u - \Delta d= g_A$, the axial coupling. 
To date, lattice calculations have
underestimated the experimental value of the axial charge, 
$g_A=1.26$, by about 20-25\%. 
The tensor charge, which is the lowest moment of the transversity function 
$h_1$\cite{JJ}, is not yet known from experiment, but may soon be measured
in polarized Drell-Yan experiments at 
Brookhaven's Relativistic Heavy Ion Collider\cite{BUNCE}. Theoretical
predictions are thus desirable. So far, Kuramashi has given the only
estimate\cite{YK}.

Another detail to note is that lattice and continuum operators
are obviously regularized in different schemes, so in any calculation
that relies on both regularizations, one must match the two, or require
them to be equal, at some renormalization scale, $\mu$. 
\begin{eqnarray}
{\cal O}(\mu)&=& Z_{\cal O}(a\mu){\cal O}^{lat}(a)
\end{eqnarray}
This matching
can be done completely in perturbation theory, but it is well known that
the bare lattice coupling constant is a poor expansion parameter and
lattice perturbation theory is more difficult, besides. An attractive
alternative is to use a non-perturbative renormalization (NPR) 
procedure\cite{GUIDO}.
The main idea is to mimic a continuum momentum (MOM) scheme on the lattice.
Recent calculations show this procedure works quite 
well\cite{DAWSON}.

\section{Domain wall fermions}

A promising new approach for treating fermions on the lattice is
domain wall fermions (DWF)\cite{KAPLAN}. DWF maintain
the full chiral symmetry of the continuum at finite lattice spacing.
Conventional lattice fermions explicitly break chiral symmetry, 
an artifact of removing the lattice
doublers, which is only restored in the continuum limit. By maintaining
this crucial continuum symmetry, simulations with DWF are more continuum-like.

The exact lattice chiral symmetry arises from the addition of an extra,
infinite, fifth dimension. The five dimensional fermions have a mass $M(s)$
in the shape of a domain wall ($M(s)<0$ for $s<0$, and $M(s)>0$ 
for $s>0$ where $s$ is the coordinate in the
fifth dimension). Four dimensional chiral zero modes 
naturally arise on the defect where $M(s)=0$.

The chiral zero modes are four dimensional in the sense that
they propagate along the domain wall in ordinary 4d space-time,
but not into the fifth dimension.
In general they have an exponentially decreasing wave function in the fifth
dimension.

For a periodic extra dimension, an anti-domain wall also appears. The
chiral zero modes on the anti-domain wall have the opposite handedness as
the ones on the domain wall. By coupling both modes to the same 4d gauge
field, we can construct a vector gauge theory, for example QCD. When the
gauge fields are explicitly four dimensional,
one can think of the extra dimension as an internal flavor space. 
This idea led to the overlap formulation of DWF\cite{NN}.

As long as the left and right handed zero modes
do not overlap in the middle of the fifth dimension,
the chiral symmetry is exact. On a finite lattice, however,
the tails overlap, and the left and right handed modes mix; 
an intrinsic quark mass is generated.
This quark mass can be made arbitrarily small by increasing the size
of the 5th dimension, $N_s$ (the number of sites in the fifth dimension). 

In practice, we use the boundary fermion variant of DWF\cite{SHAMIR}.
Here, half of the fifth dimension is discarded, so the domain walls become
the boundaries of the fifth dimension.
An explicit quark mass is introduced by coupling
the boundaries with a strength $-m$, {\it i.e.}, the chiral projections
of the five dimensional fields on the boundaries couple to each other with
strength $m$, just like a usual fermion mass term. As long the intrinsic
mass term is much smaller than $m$, it can be neglected. 

It turns out that the exponential rate of damping of 
the zero mode wave function
is controlled by the 5 dimensional quark mass, or domain wall
height $M$. Thus, there are three parameters controlling explicit chiral
symmetry breaking, $m$, $M$, and $N_s$. The formal {\it chiral} limit is
$N_s\to \infty$ and $m\to 0$.

Because they are protected by chiral symmetry, DWF have only
${\cal O}(a^2)$ errors\cite{HKN,BS}. This is true at any coupling
and in the $N_s\to\infty$ limit. At finite $N_s$ this is expected to
hold up to exponentially small corrections.

In the last section we will see an example of the power of 
DWF in QCD simulations, the calculation of the
negative-parity-nucleon/nucleon mass splitting. This calculation
also shows that DWF fermions may be a powerful tool for structure
function calculations.

\section{Review of lattice results for $g_A$.}

In this section we briefly review the current results for 
$g_A$ from lattice calculations. This is perhaps the simplest nucleon matrix
element one can imagine, and it is troubling that lattice predictions are
too low (see table~\ref{table:1}). Naively, one expects 
quenching effects to be small since only the valence quarks contribute
in the isospin limit. In fact, recent dynamical results also significantly
under predict $g_A$\cite{SESAM}.
\begin{table}[h]
   \caption{Summary of renormalized results for $\Ga$. Results are
            for Wilson fermions.       
            In cases where point-split(PS) currents were used, the
            renormalization was calculated using {\it local} currents. While
            the second entry is consistent with the measured value of 
            $\Ga=1.26$, 
            the configuration ensemble is quite small. Subscript ``$\rm con$''
            refers to the connected contribution to the matrix element.
            }
   \label{table:1}
   \begin{center}
 \begin{tabular}{|c|c|c|c|c|c|l|}\hline\hline
\hline 
type of & group & type of & lattice & $\beta$ & conf. & value \\
simulation & & current & size &  & &\\ \hline \hline
% %
% Quench & KEK\cite{KEK} 
%              & Local & $16^3 \times 20$ & 5.7 & 260 & $\Ga$=0.985(25) \\
%        &     &       &                  &     &     & $\Du$ = 0.763(35)  \\
%        &     &       &                  &     &     & $\Dd$ =-0.226(17) \\
% \hline
% %                     
% Quench &Liu et al.\cite{Liu} & Local & $16^3 \times 24$ & 6.0 & 24 
%                  & $\Ga$=1.18(11) \\       
% &  &       &   &  & & $\Za$ = 0.952 \\
% &  & PS    & $16^3 \times 24$ & 6.0 & 24& $\Ga$=1.20(10)\\
% &  &       &   &  & & $\Dq$=0.62(9) \\
% &  &       &   &  & & $\Dq^{L}$=0.65(9) \\ \hline
% %
% Quench &DESY\cite{DESY} 
%                        & PS & $16^3 \times 32$ & 6.0 & 400-1000 & $\Ga$
% =1.07(9) \\
%        &&       &   &  &    & $\Du$ = 0.938(45) \\
%        &&       &   &  &    & $\Dd$ =-0.250(12) \\ \hline 
% %
% Full ($n_{f}=2$) & SESAM\cite{SESAM} 
%                      & Local & $16^3 \times 32$ & 5.6 & 200 &
%$\Ga$=0.907(20) \\
%                  &&  &          &     &     & $\Ga^{L}$=1.243(28) \\
%                  &&  &          &     &     & $\Du$ = 0.953(25) \\
%                  &&  &          &     &     & $\Dd$ =-0.108(63) \\ \hline
%
Quench & KEK\cite{KEK} 
             & Local & $16^3 \times 20$ & 5.7 & 260 & $\Ga$=0.985(25) \\
       &     &       &                  &     &     & $\Du$ = 0.763(35)  \\
       &     &       &                  &     &     & $\Dd$ =-0.226(17) \\
       &     &       &                  &     &     & $\DS$ = 0.18(10) \\
\hline
Quench &Liu et al.\cite{Liu} & Local & $16^3 \times 24$ & 6.0 & 24 
                 & $\Ga$=1.18(11) \\       
&  & PS    & $16^3 \times 24$ & 6.0 & 24& $\Ga$=1.20(10)\\
&  &       &   &  & & $\Du$= 0.91(12) \\
&  &       &   &  & & $\Dd$=-0.30(12) \\
&  &       &   &  & & $\DS$= 0.25(12) \\ 
\hline
Quench &DESY\cite{DESY} 
                   & PS & $16^3 \times 32$ & 6.0 & 400-1000 & $\Ga$=1.07(9) \\
      &&       &   &  &    & $\Du$ = 0.830(70) \\
      &&       &   &  &    & $\Dd$ =-0.244(22) \\ 
\hline 
Full ($n_{f}=2$) & SESAM\cite{SESAM} 
                     & Local & $16^3 \times 32$ & 5.6 & 200 & $\Ga$=0.907(20) \\
                 &&  &          &     &     & $\Du$ = 0.695(18) \\ 
                 &&  &          &     &     & $\Dd$ =-0.212(8) \\
                 &&  &          &     &     & $\DS$ = 0.20(12) \\
\hline
\end{tabular}
   \end{center}
\end{table}

The problem may be related to the
renormalization of the lattice axial current $Z_A$, which in general is not 
equal to one because of explicit chiral symmetry breaking. 
The values of $\Ga$ in table~\ref{table:1} were obtained using
Wilson fermions and 
perturbative calculations of $Z_A$. These simulations are far from
the continuum limit where lattice perturbation theory is trustworthy.
There may also be significant discretization errors since each of the
studies were done at a single, rather large, lattice spacing.
 
For DWF, the conserved axial current\cite{SHAMIRandFURMAN} 
receives no renormalization. This
is not true for the local current. However, recent calculations\cite{DAWSON}
show that $Z_A/Z_V=1$, a condition required by 
the axial Ward-Takahashi identity.
Thus, a reliable estimate of $\Ga$ using DWF is given by the ratio of
the (unrenormalized) matrix elements
$\langle P|A_\mu|P\rangle/\langle P|V_\mu|P\rangle$ since $\langle
P|V_\mu|P\rangle=1$ in the continuum.

\section{Nucleon spectrum using domain wall fermions} 

\begin{figure}[h]
  \epsfxsize = 12 cm   %or \epsfysize = HEIGHT cm
      \centerline{\epsfbox{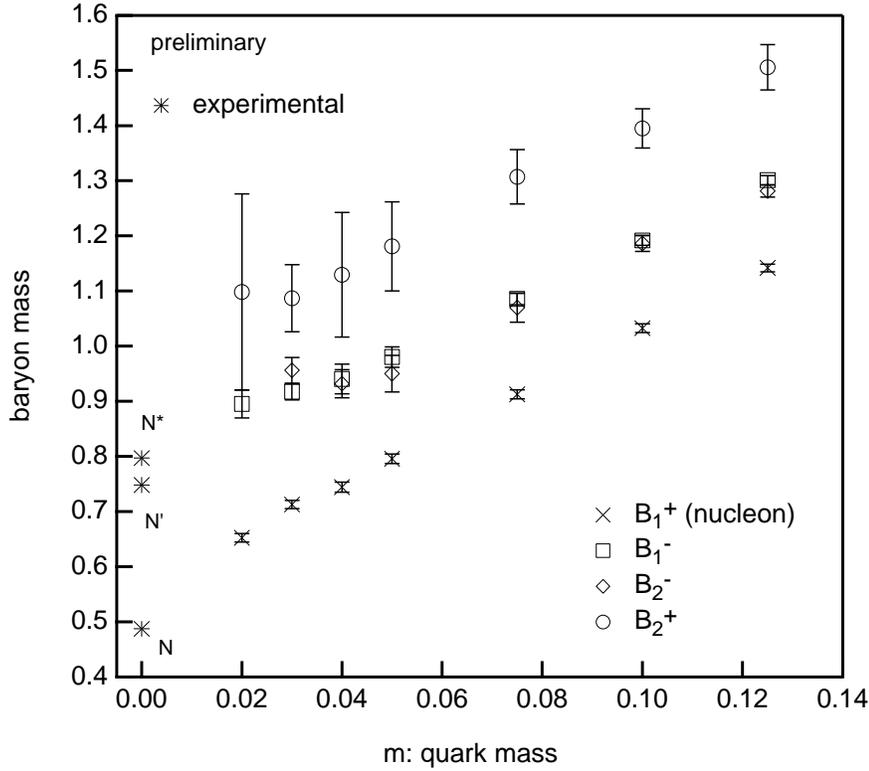}}
  \caption{The low-lying baryon spectrum in lattice units. Note the large
           mass splitting of $N$ and $N^*$ which is within 
           15\% (in the chiral limit) of the experimental value (bursts).
           The mass of the positive parity excited state is too high, however.}
  \label{fig:1}
\end{figure}

In Fig.~\ref{fig:1} we show the low energy nucleon spectrum as a 
function of the quark mass, $m$\cite{SASAKI}. The data are all for
$6/g^2=6.0$ (quenched), $M=1.8$, and lattice size $16^3\times32\times16$.
We use 205 gauge configurations for the lightest two
quark masses, $m=0.02$ and 0.03 and
24 configurations for the heavier ones, $m=0.04\to 0.125$. For an explanation
of the baryon interpolating operators $B^\pm_{1,2}$, see Ref.~\cite{SASAKI}. 
We omit the point at m=0.02 for the operator $B_2^{-}$ since
a good plateau in the effective mass plot is absent.
The experimental values for $N(939)$, $N'(1440)$, and
$N^{*}(1535)$ ($m\approx0$) (bursts) have been converted to lattice units
($a^{-1}\approx 1.9$ GeV from $am_{\rho}=0.400(8)$ in the chiral 
limit) \cite{Lingling}.

%-- ADD little bit--%
The most remarkable feature in Fig.~\ref{fig:1} is the splitting between the 
nucleon $N$ (from the $B_{1}^{+}$ operator) and its parity partner, $N^*$
(from both $B_{1}^{-}$ and $B_{2}^{-}$ operators). 
%%%%%%%%%%%%%%%%%%%%%
In the chiral limit, 
the mass difference is roughly consistent with experiment (within 15\%).
To our knowledge, this is the first such demonstration of this important
feature of the low-lying nucleon spectrum. Because of the good chiral
symmetry of DWF, this result suggests that the mass splitting is 
due to spontaneous chiral symmetry breaking. 

%-- REWRITE --%
% A remaining puzzle, however,
% is the too-high-value of the positive parity excited state. The 
% statistical errors are large for small quark mass and need to
% be improved to assess the situation.
A remaining puzzle is that we cannot extract the nucleon mass from the
unconventional nucleon operator $B_2^+$ which does not have a
non-relativistic limit.
Instead, $B_2^+$ gives a signal for the positive parity 
{\it excited} state of the nucleon. To our knowledge, 
no other groups have 
succeeded in 
extracting a clear signal for either the nucleon or its excited
state using $B_{2}^+$.
We suspect that when using DWF 
the operator $B_2^+$ has negligible overlap with the
nucleon itself, contrary to Wilson quarks 
where the operators
$B_1^+$ and $B_2^+$ mix due to explicit chiral 
symmetry breaking\cite{LEINWEBER}.
We confirm such a possibility by comparing the 
value of the $N'$ mass extracted above to the value from a
two state fit to the $B_{1}^+$ correlation function. Finally, we see no
signal for the nucleon state 
%in the $B_1^+$$B_2^+$ correlation function
in the mixed correlation function $\langle B_{1}^{+}{\bar B}_{2}^{+}+
B_{2}^{+}{\bar B}_{1}^{+}\rangle$ 
which leads us to conclude
$\langle 0|B_{2}^{+}|N\rangle\simeq 0$.

% We also note that the operator $B_2$ has neglible observed overlap with the
% nucleon, contrary to calculations with Wilson quarks where there is
% mixing bewteen $B_1$ and $B_2$ due to explicit chiral symmetry breaking.

\begin{figure}[h]
  \epsfxsize = 12 cm   %or \epsfysize = HEIGHT cm
      \centerline{\epsfbox{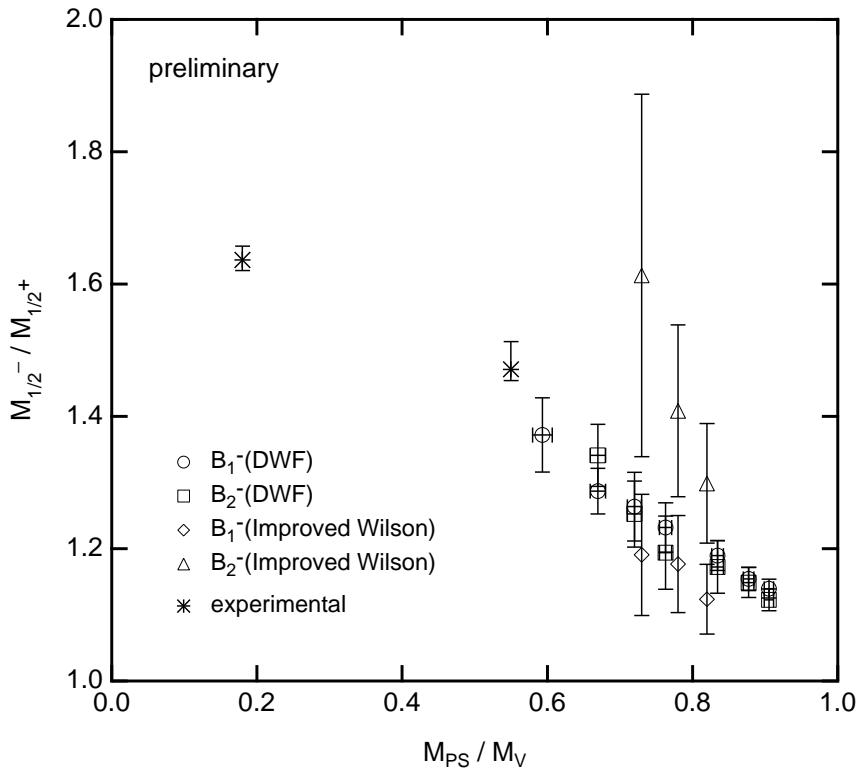}}
  \caption{Mass ratios of the low lying baryon spectrum. Experimental
      measurements are denoted by bursts. The improved
           Wilson data (diamonds and triangles) are from Ref.\cite{FXLee}. }
  \label{fig:2}
\end{figure}

In Fig.~\ref{fig:2} we show the mass ratio of the baryon parity partners
($m_{N^*}/m_N$) versus the pseudoscalar and vector meson mass ratio 
($m_\pi/m_\rho$). Experimental points are marked with bursts and
correspond to non-strange (left) and strange (right) sectors. In 
the strange sector we use $\Sigma$ and $\Sigma(1750)$ as the baryon 
parity partners and $K$ and $K^*$ for the mesons.
The baryon mass ratio grows with decreasing meson mass ratio. A naive
extrapolation is quite consistent with 
the experimental values. For comparison, we
show the recent result of Lee and Leinweber\cite{FXLee} which was obtained
using an improved Wilson quark action, D$\chi 34$, on coarse lattices.

\section{Summary}
The DWF results presented here are very encouraging. The negligible
mixing between the different operators $B_{1,2}$, and the good
agreement with the observed mass splitting of the $N$ and $N^*$ 
states indicate that DWF will work well for the more
challenging calculation of nucleon structure functions. The above
also confirms our intuition that maintaining chiral symmetry produces
more continuum-like lattice simulations.
Systematic effects due to finite volume and lattice spacing
will be addressed in future calculations. A calculation of nucleon
matrix elements using DWF is to be started soon.

\section*{Acknowledgements}
This work is part of the RIKEN-BNL-Columbia
lattice collaboration.
We are grateful to Daniel Boer, Robert Mawhinney, and Shigemi Ohta
for useful discussions.
The simulations were done on the RIKEN BNL QCDSP supercomputer. 
We thank RIKEN, Brookhaven National Laboratory, and the
U.S. Department of Energy for providing the facilities essential for
the completion of this work.

\end{document}